\documentclass[preprint]{ptephy}

\preprintnumber{KYUSHU-HET-164, KYUSHU-RCAPP-2016-03}

\usepackage{amssymb}
\usepackage{amsthm}
\usepackage{amsmath}
\usepackage{booktabs}
\usepackage{bbm}
\usepackage{mathtools}
\DeclareMathOperator{\tr}{tr}
\DeclareMathOperator{\Tr}{Tr}

\numberwithin{equation}{section}

\usepackage{hyperref}

\begin{document}

\title{Fermion number anomaly with the fluffy mirror fermion}

\author{%
\name{\fname{Ken-ichi} \surname{Okumura}}{1}
and
\name{\fname{Hiroshi} \surname{Suzuki}}{1,\ast}
}

\address{%
\affil{1}{Department of Physics, Kyushu University
744 Motooka, Nishi-ku, Fukuoka, 819-0395, Japan}
\email{hsuzuki@phys.kyushu-u.ac.jp}
}

\begin{abstract}
Quite recently, Grabowska and Kaplan presented a 4-dimensional lattice
formulation of chiral gauge theories based on the chiral overlap operator. We
study this formulation from the perspective of the fermion number anomaly and
possible associated phenomenology. A simple argument shows that the consistency
of the formulation implies that the fermion with the opposite chirality to the
physical one, the ``fluffy mirror fermion'' or ``fluff'', suffers from the
fermion number anomaly in the same magnitude (with the opposite sign) as the
physical fermion. This immediately shows that if at least one of the fluff
quarks is massless, the formulation provides a simple viable solution to the
strong CP problem. Also, if the fluff interacts with gravity essentially in the
same way as the physical fermion, the formulation can realize the asymmetric
dark matter scenario.
\end{abstract}
\subjectindex{B01, B31, B46, B70}
\maketitle

\section{Introduction}
\label{sec:1}
Quite recently~\cite{Grabowska:2016,Kaplan:2016}, Grabowska and Kaplan
presented a 4-dimensional lattice formulation of chiral gauge theories based
on the chiral overlap operator. They obtained this 4-dimensional formulation by
taking the infinite 5th-dimensional extent limit in their 5-dimensional
domain-wall formulation in~Ref.~\cite{Grabowska:2015qpk}.\footnote{For a
closely related 6-dimensional domain-wall formulation,
see~Ref.~\cite{Fukaya:2016ofi}.} The procedure is analogous to the one obtained
by the overlap formulation~\cite{Narayanan:1992wx,Narayanan:1993sk,%
Narayanan:1993ss,Narayanan:1994gw} and the associated overlap Dirac
operator~\cite{Neuberger:1997fp,Neuberger:1998wv} from the domain-wall Dirac
fermion~\cite{Kaplan:1992bt,Shamir:1993zy,Furman:1994ky}. Although the work
of~Refs.~\cite{Luscher:1998du,Luscher:1999un} already paved the way to a
manifestly gauge-invariant nonperturbative formulation of chiral gauge
theories, the construction of the gauge-invariant fermion
measure~\cite{Luscher:1998du,Luscher:1999un} still remains elusive, except for
the abelian case~\cite{Luscher:1998du} for which the explicit construction is
known. See also~Refs.~\cite{Kikukawa:2000kd,Kikukawa:2001mw,Kadoh:2007xb}.
Thus, the manifestly gauge-invariant lattice formulation
in~Refs.~\cite{Grabowska:2016,Kaplan:2016} appears astonishing in our opinion,
because it can avoid the complicated construction of the fermion measure;
however, one still has to understand how the locality of the formulation
distinguishes the anomaly-free cases from the anomalous cases.

In the present note, we study the formulation
in~Refs.~\cite{Grabowska:2016,Kaplan:2016} from the perspective of the fermion
number anomaly~\cite{'tHooft:1976up,Fujikawa:1983bg} and possible associated
phenomenology. A simple argument shows that the consistency of the formulation
implies that the fermion with the opposite chirality to the physical one, the
``fluffy mirror fermion'' or ``fluff'', suffers from the fermion number anomaly
in the same magnitude (with the opposite sign) as the physical chiral fermion;
the fluff feels the same topological charge as the physical one, contrary to
the claim made in~Refs.~\cite{Grabowska:2016,Kaplan:2016}. This must be so
simply because the number symmetry that rotates both the physical and the fluff
fermions is an exact symmetry of the formulation. Assuming the validity of the
formulation, especially the restoration of the locality for anomaly-free chiral
gauge theories, we accept the existence of the fluff fermions positively and
discuss possible phenomenological implications. The above property of the
fermion number anomaly immediately shows that if at least one of the fluff
quarks is massless, the formulation provides a simple viable solution to the
strong CP problem; this is the massless up-quark solution in the mirror or
fluff sector. Also, if the fluff interacts with gravity essentially in the same
way as the physical fermions, its fermion number violation can provide an
understanding of the coincidence between the baryon and the dark matter
abundances, realizing the asymmetric dark matter
scenario~\cite{Kitano:2004sv,Kaplan:2009ag}.

\section{Fermion number anomaly with the chiral overlap operator}
\label{sec:2}
In the formulation of~Refs.~\cite{Grabowska:2016,Kaplan:2016}, the partition
function of the (left-handed) Weyl fermion is given by\footnote{In what
follows, the lattice gauge field is treated as a nondynamical background,
because the integration over the gauge field is irrelevant in the following
discussion.}
\begin{equation}
   \int\mathcal{D}\psi\mathcal{D}\Bar{\psi}\,
   \exp\left(-a^4\sum_x\Bar{\psi}(x)\mathcal{D}_\chi\psi(x)\right),
\label{eq:(2.1)}
\end{equation}
where $a$ denotes the lattice spacing and $\mathcal{D}_\chi$ is the chiral
overlap operator defined by
\begin{equation}
   a\mathcal{D}_\chi
   =1+\gamma_5\left[1-(1-\epsilon_\star)\frac{1}{\epsilon\epsilon_\star+1}
   (1-\epsilon)\right].
\label{eq:(2.2)}
\end{equation}
Here, $\epsilon$ is the sign function 
\begin{equation}
   \epsilon=\epsilon(H_w)=\frac{H_w}{\sqrt{H_w^2}}
\label{eq:(2.3)}
\end{equation}
of the hermitian Wilson Dirac operator
\begin{equation}
   H_w=\gamma_5\left[\frac{1}{2}\gamma_\mu(\nabla_\mu+\nabla_\mu^*)
   -\frac{1}{2}a\nabla_\mu\nabla_\mu^*-m\right],
\label{eq:(2.4)}
\end{equation}
where $\nabla_\mu$ and~$\nabla_\mu^*$ are forward and backward gauge-covariant
lattice derivatives, respectively; $H_w$ is a function of the gauge field~$A$.
The gauge field contained in~$\epsilon_\star$, denoted by $A_\star$, is on the
other hand defined by a one-parameter deformation (the gradient flow) of the
original 4-dimensional gauge field~$A$~\cite{Grabowska:2016,Kaplan:2016}; i.e.,
\begin{equation}
   \epsilon=\epsilon(H_w[A]),\qquad\epsilon_\star=\epsilon(H_w[A_\star]).
\label{eq:(2.5)}
\end{equation}
The crucial point in the formulation is that the fermion integration variables
in~Eq.~\eqref{eq:(2.1)} are 4-component Dirac fields without any chirality
constraint. This is quite in contrast to the formulation
of~Refs.~\cite{Luscher:1998du,Luscher:1999un} and for this reason one can avoid
the construction of the nontrivial fermion measure. The naive continuum limit
of Eq.~\eqref{eq:(2.2)} is~\cite{Grabowska:2016,Kaplan:2016}
\begin{equation}
   \mathcal{D}_\chi\stackrel{a\to0}{\to}
   \frac{1}{am}\begin{pmatrix}0&\sigma_\mu D_\mu(A)\\
   \Bar{\sigma}_\mu D_\mu(A_\star)&0\\
   \end{pmatrix}.
\label{eq:(2.6)}
\end{equation}
Therefore, in this naive continuum limit, the left-handed fermion interacts
with~$A$ and the right-handed one interacts with~$A_\star$. If the gauge
field~$A_\star$ is basically trivial, then the system may be regarded as the
left-handed Weyl fermion interacting with the gauge
field~\cite{AlvarezGaume:1983cs}. In~Refs.~\cite{Grabowska:2016,Kaplan:2016},
$A_\star$ is defined by the gradient flow~\cite{Narayanan:2006rf,%
Luscher:2009eq,Luscher:2010iy,Luscher:2011bx} along the 5th-dimensional
direction of the domain-wall formulation in~Ref.~\cite{Grabowska:2015qpk}.

Now, using $\epsilon^2=\epsilon_\star^2=1$,
\begin{equation}
   \epsilon+\epsilon_\star=\epsilon(1+\epsilon\epsilon_\star)
\label{eq:(2.7)}
\end{equation}
and one can confirm that
\begin{equation}
   \left[1-(1-\epsilon_\star)\frac{1}{\epsilon\epsilon_\star+1}
   (1-\epsilon)\right]^2=1
\label{eq:(2.8)}
\end{equation}
and, consequently, $\mathcal{D}_\chi$ satisfies the Ginsparg--Wilson
relation~\cite{Ginsparg:1981bj}
\begin{equation}
   \gamma_5\mathcal{D}_\chi+\mathcal{D}_\chi\gamma_5
   =a\mathcal{D}_\chi\gamma_5\mathcal{D}_\chi.
\label{eq:(2.9)}
\end{equation}
This Ginsparg--Wilson relation allows one to define a modified
$\gamma_5$~\cite{Luscher:1998pqa,Niedermayer:1998bi},
\begin{equation}
   \Hat{\gamma}_5\equiv\gamma_5(1-a\mathcal{D}_\chi),
\label{eq:(2.10)}
\end{equation}
which satisfies\footnote{Note that in general
$(\Hat{\gamma}_5)^\dagger\neq\Hat{\gamma}_5$. We would like to thank Okuto
Morikawa for pointing this out to us.}
\begin{equation}
   (\Hat{\gamma}_5)^2=1,\qquad
   \mathcal{D}_\chi\Hat{\gamma}_5=-\gamma_5\mathcal{D}_\chi.
\label{eq:(2.11)}
\end{equation}
Thus, we can define projection operators,
\begin{equation}
   \Hat{P}_\pm\equiv\frac{1}{2}(1\pm\Hat{\gamma}_5),\qquad
   P_\pm\equiv\frac{1}{2}(1\pm\gamma_5),
\label{eq:(2.12)}
\end{equation}
and chiral components of the fermion by
\begin{align}
   \Hat{P}_-\psi_L(x)&=\psi_L(x),&\Bar{\psi}_L(x)P_+=\Bar{\psi}_L(x),
\label{eq:(2.13)}
\\
   \Hat{P}_+\psi_R(x)&=\psi_R(x),&\Bar{\psi}_R(x)P_-=\Bar{\psi}_R(x).
\label{eq:(2.14)}
\end{align}
Then, thanks to the last relation of~Eq.~\eqref{eq:(2.11)}, the action is
completely decomposed into the left and right fermion components as
\begin{equation}
   a^4\sum_x\Bar{\psi}(x)\mathcal{D}_\chi\psi(x)
   =a^4\sum_x\left[\Bar{\psi}_L(x)\mathcal{D}_\chi\psi_L(x)
   +\Bar{\psi}_R(x)\mathcal{D}_\chi\psi_R(x)\right].
\label{eq:(2.15)}
\end{equation}

Let us now introduce the fermion number transformations. For the left-handed
``physical'' fermion, the fermion number~$U(1)$ transformation is defined by
\begin{equation}
   \psi_L(x)\to e^{i\theta}\psi_L(x),\qquad
   \Bar{\psi}_L(x)\to e^{-i\theta}\Bar{\psi}_L(x),
\label{eq:(2.16)}
\end{equation}
and $\psi_R(x)$ and~$\Bar{\psi}_R(x)$ are kept intact. On the other hand, for
the right-handed ``invisible'' or ``fluff'' fermion,
\begin{equation}
   \psi_R(x)\to e^{i\theta}\psi_R(x),\qquad
   \Bar{\psi}_R(x)\to e^{-i\theta}\Bar{\psi}_R(x),
\label{eq:(2.17)}
\end{equation}
and $\psi_L(x)$ and~$\Bar{\psi}_L(x)$ are kept intact. We note that the
combination of these two, which defines the sum of fermion numbers of both
chiralities,
\begin{align}
   \psi_L(x)&\to e^{i\theta}\psi_L(x),&
   \Bar{\psi}_L(x)&\to e^{-i\theta}\Bar{\psi}_L(x),
\label{eq:(2.18)}
\\
   \psi_R(x)&\to e^{i\theta}\psi_R(x),&
   \Bar{\psi}_R(x)&\to e^{-i\theta}\Bar{\psi}_R(x),
\label{eq:(2.19)}
\end{align}
or, equivalently,
\begin{equation}
   \psi(x)\to e^{i\theta}\psi(x),\qquad
   \Bar{\psi}(x)\to e^{-i\theta}\Bar{\psi}(x),   
\label{eq:(2.20)}
\end{equation}
leaves the action and the fermion measure in~Eq.~\eqref{eq:(2.1)} invariant.
Therefore, this symmetry is exact and anomaly-free.

The fermion number of the left-handed physical fermion is anomalous. The
action~\eqref{eq:(2.15)} is invariant under~Eq.~\eqref{eq:(2.16)}, but the
measure is not because of the chirality projection~\eqref{eq:(2.13)} which
nontrivially depends on the gauge field. Considering the
transformation~\eqref{eq:(2.16)} with the localized parameter
$\theta\to\theta(x)$, we have the anomalous conservation law of the fermion
number current of the left-handed fermion, 
\begin{align}
   \left\langle\partial_\mu j_{L\mu}(x)\right\rangle
   &=\tr\left[\Hat{P}_-(x,x)-P_+\delta_{x,x}\right]
\notag\\
   &=\frac{1}{2}\tr\left[\gamma_5 a\mathcal{D}_\chi(x,x)\right],
\label{eq:(2.21)}
\end{align}
where $\tr$ stands for the trace over the spinor and gauge indices only and
$\Hat{P}_-(x,y)$, e.g., denotes the kernel of~$\Hat{P}_-$ in the position
space; we have used $\tr\gamma_5=0$ in the second equality. Similarly, the
fermion number current of the right-handed fermion does not conserve:
\begin{align}
   \left\langle\partial_\mu j_{R\mu}(x)\right\rangle
   &=\tr\left[\Hat{P}_+(x,x)-P_-\delta_{x,x}\right]
\notag\\
   &=-\frac{1}{2}\tr\left[\gamma_5 a\mathcal{D}_\chi(x,x)\right].
\label{eq:(2.22)}
\end{align}
The sum of these two conserves
\begin{equation}
   \left\langle
   \partial_\mu\left[j_{L\mu}(x)+j_{R\mu}(x)\right]\right\rangle=0
\label{eq:(2.23)}
\end{equation}
as should be from the invariance of the system under Eq.~\eqref{eq:(2.20)}.
Already Eqs.~\eqref{eq:(2.21)} and~\eqref{eq:(2.22)} show that the left-handed
physical fermion and its mirror partner, the right-handed fluff, suffer from
the fermion number anomaly in the same magnitude (with the opposite sign), no
matter how $A$ and~$A_\star$ are related. This is contrary to the claims
of~Refs.~\cite{Grabowska:2016,Kaplan:2016} that the flowed field~$A_\star$,
which couples to the right-handed fermion, is pure-gauge and thus carries no
topological information about the original gauge field~$A$.

That $A$ and~$A_\star$ must carry the same topological information can also be
seen from the nonconservation of the fermion number charge, which is obtained
by integrating Eq.~\eqref{eq:(2.21)} over 4-dimensional spacetime. Using
Eq.~\eqref{eq:(2.2)}, we have
\begin{align}
   \left\langle Q_L(x_0=+\infty)\right\rangle
   -\left\langle Q_L(x_0=-\infty)\right\rangle
   &=\frac{1}{2}\Tr
   \left[1-(1-\epsilon_\star)\frac{1}{\epsilon\epsilon_\star+1}
   (1-\epsilon)\right]
\notag\\
   &=\frac{1}{2}\Tr
   \left[1-(1-\epsilon)(1-\epsilon_\star)\frac{1}{\epsilon\epsilon_\star+1}
   \right]
\notag\\
   &=\frac{1}{2}\Tr\left[
   (\epsilon+\epsilon_\star)\frac{1}{\epsilon\epsilon_\star+1}\right]
\notag\\
   &=\frac{1}{2}\Tr\epsilon.
\label{eq:(2.24)}
\end{align}
In this expression, $\Tr\equiv a^4\sum_x\tr$ and we have used the cyclic
property of~$\Tr$ in the first equality. In the last equality, we have
noted~Eq.~\eqref{eq:(2.7)}. However, we can equally use
$\epsilon+\epsilon_\star=(1+\epsilon\epsilon_\star)\epsilon_\star$ to conclude
\begin{equation}
   \left\langle Q_L(x_0=+\infty)\right\rangle
   -\left\langle Q_L(x_0=-\infty)\right\rangle
   =\frac{1}{2}\Tr\epsilon_\star.
\label{eq:(2.25)}
\end{equation}
Thus, for the expression~\eqref{eq:(2.2)} to be meaningful,
$\Tr\epsilon_\star=\Tr\epsilon$. In particular, $A_\star$ cannot be pure gauge,
for which we can see that $\Tr\epsilon_\star=0$.

The above argument implicitly assumes that $\Tr\epsilon$ can be nonzero. This
is actually the case as can be seen in the classical continuum
limit~\cite{Kikukawa:1998pd,Fujikawa:1998if,Adams:1998eg,Suzuki:1998yz},
\begin{align}
   \frac{1}{2}\tr\epsilon(x,x)
   &=\frac{1}{2}\tr\frac{H_w}{\sqrt{H_w^2}}(x,x)
\notag\\
   &\stackrel{a\to0}{\to}
   -\frac{1}{32\pi^2}I(am,1)\epsilon_{\mu\nu\rho\sigma}
   \tr\left[F_{\mu\nu}(x)F_{\rho\sigma}(x)\right]
\label{eq:(2.26)}
\end{align}
for a smooth background gauge field, where the function~$I(am,r)$ is given by
\begin{equation}
   I(am,r)=\theta(am/r)-4\theta(am/r-2)+6\theta(am/r-4)-4\theta(am/r-6)
   +\theta(am/r-8)
\label{eq:(2.27)}
\end{equation}
from the step function~$\theta(x)$. For~$0<am<2$ for which the free lattice
Dirac operator possesses only a single zero in the Brillouin zone, $I(am,1)=1$
and Eq.~\eqref{eq:(2.26)} reproduces the correct fermion number anomaly in the
classical continuum limit.\footnote{We note that the topological charge is
given by~$Q_{\text{top.}}=\int d^4x\,(1/64\pi^2)\epsilon_{\mu\nu\rho\sigma}
F_{\mu\nu}^a(x)F_{\rho\sigma}^a(x)$ (for the instanton, $Q_{\text{top.}}=1$)
and thus in the continuum limit,
$\langle Q_L(x_0=+\infty)\rangle-\langle Q_L(x_0=-\infty)\rangle=%
2N_fT(R)Q_{\text{top.}}$, where $N_f$ is the number of flavors
and~$\tr(T^aT^b)=-T(R)\delta^{ab}$.} Thus,
$\Tr\epsilon=a^4\sum_x\tr\epsilon(x,x)$ can actually be nonzero for a
topologically nontrivial background.

It is of interest to obtain the classical continuum limit of the right-hand
side of~Eq.~\eqref{eq:(2.21)} which depends on both $A$ and~$A_\star$:
\begin{equation}
   \mathcal{A}_L(x)\equiv\frac{1}{2}\tr\left[1-(1-\epsilon_\star)
   \frac{1}{\epsilon\epsilon_\star+1}(1-\epsilon)\right](x,x).
\label{eq:(2.28)}
\end{equation}
We could not complete the explicit perturbative calculation, but a general
argument suggests its form in the classical continuum limit as follows. First,
since $\epsilon$ and~$\epsilon_\star$ are hermitian operators,
$\mathcal{A}_L(x)[A_\star,A]=\mathcal{A}_L(x)[A,A_\star]^*$. Next,
from~Eqs.~\eqref{eq:(2.24)}, \eqref{eq:(2.25)}, and~\eqref{eq:(2.26)},
\begin{align}
   \mathcal{A}_L(x)
   &\stackrel{a\to0}{\to}-\frac{1}{32\pi^2}\epsilon_{\mu\nu\rho\sigma}
   \tr\left[F_{\mu\nu}(x)F_{\rho\sigma}(x)\right]+\partial_\mu X_\mu(x),
\label{eq:(2.29)}
\\
   \mathcal{A}_L(x)
   &\stackrel{a\to0}{\to}-\frac{1}{32\pi^2}\epsilon_{\mu\nu\rho\sigma}
   \tr\left[F_{\star\mu\nu}(x)F_{\star\rho\sigma}(x)\right]+\partial_\mu Y_\mu(x),
\label{eq:(2.30)}
\end{align}
where $X_\mu(x)$ and~$Y_\mu(x)$ are gauge-invariant vectors. From these
properties, we conjecture that the classical continuum limit
of~$\mathcal{A}_L(x)$ is given by
\begin{align}
   \mathcal{A}_L(x)&\stackrel{a\to0}{\to}
   -\frac{1}{64\pi^2}\left\{
   \epsilon_{\mu\nu\rho\sigma}
   \tr\left[F_{\mu\nu}(x)F_{\rho\sigma}(x)\right]
   +\epsilon_{\mu\nu\rho\sigma}
   \tr\left[F_{\star\mu\nu}(x)F_{\star\rho\sigma}(x)\right]
   \right\}.
\label{eq:(2.31)}
\end{align}
This is obviously consistent with the
property~$\mathcal{A}_L(x)[A_\star,A]=\mathcal{A}_L(x)[A,A_\star]^*$. Also
noting the identity,
\begin{align}
   &\epsilon_{\mu\nu\rho\sigma}
   \tr\left[F_{\star\mu\nu}(x)F_{\star\rho\sigma}(x)\right]
\notag\\
   &=\epsilon_{\mu\nu\rho\sigma}
   \tr\left[F_{\mu\nu}(x)F_{\rho\sigma}(x)\right]
   +\partial_\mu\int_0^\infty dt\,4
   \epsilon_{\mu\nu\rho\sigma}
   \tr\left[\frac{\partial}{\partial t}A_\nu(t,x)
   F_{\rho\sigma}(t,x)\right],
\label{eq:(2.32)}
\end{align}
where $A_\mu(t,x)$ is a one-parameter family of the gauge fields that connects
$A$ and~$A_\star$,
\begin{equation}
   A_\mu(t=0,x)=A_\mu(x),\qquad A_\mu(t=\infty,x)=A_{\star\mu}(x),
\label{eq:(2.33)}
\end{equation}
Eqs.~\eqref{eq:(2.29)} and~\eqref{eq:(2.30)} are satisfied if
$\frac{\partial}{\partial t}A_\nu(t,x)$ is gauge covariant. The simplest
possibility for this would be the gradient flow
\begin{equation}
   \frac{\partial}{\partial t}A_\mu(t,x)
   =D_\nu F_{\nu\mu}(t,x).
\label{eq:(2.34)}
\end{equation}

\section{Phenomenological/cosmological implications}
\label{sec:3}
In this section, we discuss possible phenomenological/cosmological implications
of the lattice formulation in~Refs.~\cite{Grabowska:2016,Kaplan:2016}. We have
observed that, in this lattice formulation of the chiral gauge theory, the
fermion number violation caused by the chiral gauge interaction for the
``physical'' or ``visible'' fermion necessarily accompanies the opposite
fermion number violation for the opposite chirality ``invisible'' fermion, the
``fluff''. If the standard model~(SM) is nonperturbatively defined by the
present lattice formulation (``fluffy SM''), all the SM fermions accompany
their mirror partners. This fact immediately implies that if at least one of
the fluff quarks is massless, the $\theta$~angle for the strong interaction can
be rotated away by the chiral rotation of the fluff. This is simply the
massless up-quark solution to the strong CP problem in the fluff or mirror
sector. Note that the structure of Yukawa couplings in the fluff sector (see
below) can be quite different from the physical sector.

The properties of the formulation also possibly have implications for the
baryogenesis and the dark matter problem. The $B+L$ charge violation that may
be caused by the instanton or sphaleron effect in the electroweak sector always
accompanies a $B+L$ violation with the same magnitude (with an opposite sign)
for the fluff; the sum $B+L+(B+L)_{\text{fluff}}$ is conserved. The $B-L$ charges
in the visible sector and those in the fluff sector are, on the other hand,
separately conserved. These properties provide a realization of the asymmetric
dark matter scenario~\cite{Nussinov:1985xr,Barr:1990ca,Kaplan:1991ah,%
Kuzmin:1996he,Hooper:2004dc,Kitano:2004sv,Kaplan:2009ag} as follows.

First, we assume that the left-handed physical (or anti-right-handed) fermions
and the right-handed fluff interact with gravity essentially in the same way,
i.e., the fermion is coupled to the gravity as an ordinary Dirac fermion, as
naturally suggested from the action~\eqref{eq:(2.15)} in flat
spacetime.\footnote{Such a left--right symmetric gravitational coupling will
imply the equality (with the opposite sign) of the gravitational contributions
to the fermion number anomaly between the physical fermion and the fluff. It
will also be interesting to investigate the possible implications of the
gravitational interaction between the physical and fluff sectors.}

Suppose further that the fluff is thermalized after the inflation, e.g., via a
coupling with the inflaton. Then it will contribute to the relativistic degrees
of freedom other than the photon, $N_{\text{eff}}$, at the epoch of the
recombination. $N_{\text{eff}}$ is constrained from the observation of the
cosmic microwave background as~$N_{\text{eff}}<3.13\pm0.32$~\cite{Ade:2015xua}.
Therefore, it is plausible that the fluff acquires mass and annihilates into
the visible particles (or to a dark radiation if we allow $N_{\text{eff}}=4$)
via interactions other than those from the SM gauge group. For example, we can
couple a Higgs field to the fluff to make them heavy,\footnote{To make the
``fluff neutrino'' massive, we may turn off the seesaw mechanism by
eliminating the Majorana neutrino mass if $B-L$ is global, while we tune the
neutrino Yukawa couplings if $B-L$ is gauged.}
\begin{equation}
   y\Bar{u}_LHq_R+\text{h.c.},
\label{eq:(3.1)}
\end{equation}
where $u_L$ denotes the ``fluff up quark'' and $q_R$ the ``fluff quark
doublet''. To make the annihilation process efficient, we may further introduce
a light scalar field that couples to the fluffs with nonhierarchical Yukawa
couplings; another possibility is to introduce a light $B-L$ gauge boson that
couples to the fluffs by the minimal coupling and to the physical fermions
through the gradient flow.

Once the baryon or lepton number is produced in either side (e.g., as in
electroweak baryogenesis~\cite{Kuzmin:1985mm,Shaposhnikov:1986jp,%
Shaposhnikov:1987tw,Cohen:1993nk} or leptogenesis~\cite{Fukugita:1986hr}) when
the sphaleron process is at work, it will be redistributed between the two
sectors~\cite{Barr:1990ca,Kaplan:1991ah}.\footnote{This scenario is essentially
the fluff version of~Ref.~\cite{Kaplan:1991ah}
with~$X=(B+L)_{\text{fluff}}$.} Thus the two sectors obtain a similar size of
baryon and lepton numbers. Assuming the SM gauge group, $N_g$~generations and
$N_h$~Higgs doublets, the conserved changes are Cartan generators of~$SU(3)_c$
and~$SU(2)_L$, the $U(1)_Y$ hypercharge, the difference between the baryon and
lepton numbers in the physical sector~$B-L$, that in the fluff
sector~$(B-L)_{\text{fluff}},$\footnote{Here we consider the case in which the
$B-L$ breaking effects are decoupled.} and the sum of the baryon and lepton
numbers $B+L+(B+L)_{\text{fluff}}$. Then the standard
argument~\cite{Harvey:1990qw,Weinberg:2008zzc} on the basis of the relativistic
free particle approximation yields, at the chemical equilibrium,
\begin{align}
   n_B&=\frac{21}{52}n_{B-L}+\frac{1}{4}n_{B+L+(B+L)_{\text{fluff}}}
   +\frac{5}{52}n_{(B-L)_{\text{fluff}}},
\label{eq:(3.2)}
\\
   n_L&=-\frac{31}{52}n_{B-L}+\frac{1}{4}n_{B+L+(B+L)_{\text{fluff}}}
   +\frac{5}{52}n_{(B-L)_{\text{fluff}}},
\label{eq:(3.3)}
\\
   n_{B_{\text{fluff}}}&=\frac{5}{52}n_{B-L}+\frac{1}{4}n_{B+L+(B+L)_{\text{fluff}}}
   +\frac{21}{52}n_{(B-L)_{\text{fluff}}},
\label{eq:(3.4)}
\\
   n_{L_{\text{fluff}}}&=\frac{5}{52}n_{B-L}+\frac{1}{4}n_{B+L+(B+L)_{\text{fluff}}}
   -\frac{31}{52}n_{(B-L)_{\text{fluff}}}.
\label{eq:(3.5)}
\end{align}
For example, for the standard leptogenesis for which initially $n_L\neq0$ and
$n_B=n_{B_{\text{fluff}}}=n_{L_{\text{fluff}}}=0$, we have
$n_{B_{\text{fluff}}}=n_{L_{\text{fluff}}}=-n_B$ at the equilibrium. It is
interesting that in the present formulation,
Eqs.~\eqref{eq:(3.2)}--\eqref{eq:(3.5)} depend on neither~$N_g$ nor~$N_h$.
Also, note that $B+L$ produced in the early universe is not washed out and
split into the two sectors. $n_{B_\text{fluff}}$ and~$n_{L_\text{fluff}}$ are fixed
after the universe is cooled down below the temperature at which the sphaleron
process is effective. Then if pair-annihilation processes are efficient enough,
the symmetric part of the fluff sector disappears. The remaining fluff fermion
(or minus anti-fermion) numbers are proportional to~$n_{B_\text{fluff}}$
and~$n_{L_\text{fluff}}$ and provide the observed dark matter abundance
$\Omega_{\text{DM}}\simeq5\Omega_B$ if the average of fluff fermion masses par
baryon or lepton number is~$O(5)\,\text{GeV}$.\footnote{If masses of the fluff
fermions are provided by the Yukawa coupling to the SM Higgs field, the Yukawa
couplings to the fluff fermions are constrained by the upper bound on the Higgs
invisible branching ratio at LHC. The fluff mass of~$O(1)\,\text{GeV}$ is
barely consistent with the present bound~\cite{Aad:2015pla,CMS:2016rfr} and the
case of the averaged fluff fermion mass of~$O(5)\,\text{GeV}$ might be
excluded. This number however can be reduced within the present scenario by
assuming that $n_B\neq0$ and~$n_L\neq0$ for the primordial densities.}
\footnote{In the minimal setting, the flavor is preserved in the ``invisible''
sector since the weak interaction is effectively suppressed by the gradient
flow.} It is interesting to observe that the SM $SU(3)$ gauge interaction does
not contribute to the renormalization group running of the Yukawa couplings of
the fluff quarks. This would reduce the ratio of the ``fluff top'' Yukawa
coupling and the physical top Yukawa coupling even if they are equal at the
cutoff scale.

Another interesting observation is that the visible quarks do not contribute to
the neutron electric dipole moment (EDM) after the strong CP phase is rotated
away to the fluff quarks. Thus the stringent bound on the strong CP-violating
phase could be ameliorated even if all the fluff quarks are massive. It is
tempting to imagine the phase participates in the baryo/leptogenesis. Actually,
the fluff quarks couple to the neutron EDM via the vacuum polarization of the
physical Higgs or the $B-L$ gauge field. It is at least one-loop suppressed. In
addition, the diagram is suppressed by the two small visible Yukawa couplings
or flowed $B-L$ couplings. Also, the CP-violating fluff--Higgs coupling is
suppressed by the smallest fluff Yukawa coupling according to the standard
calculation of the EDM~\cite{Crewther:1979pi}.

The above fluff dark matter interacts with the SM particle via the Higgs
portal~\cite{Patt:2006fw,Kim:2006af,MarchRussell:2008yu,Kim:2008pp,%
Kanemura:2010sh,Djouadi:2011aa,Low:2011kp,Englert:2011aa} or a $B-L$ gauge
field~\cite{Mohapatra:1980qe,Wetterich:1981bx,Buchmuller:1991ce,Khalil:2006yi,%
Basso:2008iv}. They can be explored or constrained by direct/indirect or
collider dark matter searches. In the former case, they also contribute to the
invisible decay width of the SM Higgs boson. In the latter case, the $B-L$
gauge boson will be searched by the LHC and future colliders. The dark matter
is multicomponent in the minimal setup and its precise nature depends on the
hierarchy of the fluff Yukawa couplings.

We have briefly sketched phenomenological/cosmological implications of the
formulation. The construction of working phenomenological models and their
detailed numerical analyses are beyond the scope of this paper; we leave them
to the future work. It should be noted that the present lattice formulation, in
principle, allows one to carry out nonperturbative field theoretical analyses
of, e.g., the electroweak baryogenesis in the fluffy SM, from first principles.

\section*{Note added}
An explicit calculation~\cite{Makino:2016auf} of the classical continuum limit
of the fermion number anomaly~\eqref{eq:(2.28)} reveals that the conjectured
form~\eqref{eq:(2.31)} is wrong; the correct expression turns out to be much
more complicated as shown in~Ref.~\cite{Makino:2016auf}.

\section*{Acknowledgements}
We would like to thank
Hidenori Fukaya,
Dorota M. Grabowska,
David B. Kaplan,
Tetsuya Onogi,
and
Ryo Yamamura
for explanations of their works.
The work of H.~S. is supported in part by JSPS Grants-in-Aid for Scientific
Research Grant Number~JP16H03982.


\begin{thebibliography}{00}

\bibitem{Grabowska:2016}
D.~M.~Grabowska,
``Continuing the saga of fluffy mirror fermions,''
talk delivered at the 34th International Symposium on Lattice Field Theory,
\url{https://conference.ippp.dur.ac.uk/event/470/session/16/contribution/364}

\bibitem{Kaplan:2016}
D.~B.~Kaplan,
``A new perspective on chiral gauge theories,''
talk delivered at the 34th International Symposium on Lattice Field Theory,
\url{https://conference.ippp.dur.ac.uk/event/470/session/1/contribution/398}

\bibitem{Grabowska:2015qpk} 
  D.~M.~Grabowska and D.~B.~Kaplan,
  Phys.\ Rev.\ Lett.\  {\bf 116}, no. 21, 211602 (2016)
  doi:10.1103/PhysRevLett.116.211602
  [arXiv:1511.03649 [hep-lat]].

\bibitem{Fukaya:2016ofi} 
  H.~Fukaya, T.~Onogi, S.~Yamamoto and R.~Yamamura,
  arXiv:1607.06174 [hep-th].

\bibitem{Narayanan:1992wx} 
  R.~Narayanan and H.~Neuberger,
  Phys.\ Lett.\ B {\bf 302}, 62 (1993)
  doi:10.1016/0370-2693(93)90636-V
  [hep-lat/9212019].

\bibitem{Narayanan:1993sk} 
  R.~Narayanan and H.~Neuberger,
  Nucl.\ Phys.\ B {\bf 412}, 574 (1994)
  doi:10.1016/0550-3213(94)90393-X
  [hep-lat/9307006].

\bibitem{Narayanan:1993ss} 
  R.~Narayanan and H.~Neuberger,
  Phys.\ Rev.\ Lett.\  {\bf 71}, no. 20, 3251 (1993)
  doi:10.1103/PhysRevLett.71.3251
  [hep-lat/9308011].

\bibitem{Narayanan:1994gw} 
  R.~Narayanan and H.~Neuberger,
  Nucl.\ Phys.\ B {\bf 443}, 305 (1995)
  doi:10.1016/0550-3213(95)00111-5
  [hep-th/9411108].

\bibitem{Neuberger:1997fp} 
  H.~Neuberger,
  Phys.\ Lett.\ B {\bf 417}, 141 (1998)
  doi:10.1016/S0370-2693(97)01368-3
  [hep-lat/9707022].

\bibitem{Neuberger:1998wv} 
  H.~Neuberger,
  Phys.\ Lett.\ B {\bf 427}, 353 (1998)
  doi:10.1016/S0370-2693(98)00355-4
  [hep-lat/9801031].

\bibitem{Kaplan:1992bt} 
  D.~B.~Kaplan,
  Phys.\ Lett.\ B {\bf 288}, 342 (1992)
  doi:10.1016/0370-2693(92)91112-M
  [hep-lat/9206013].

\bibitem{Shamir:1993zy} 
  Y.~Shamir,
  Nucl.\ Phys.\ B {\bf 406}, 90 (1993)
  doi:10.1016/0550-3213(93)90162-I
  [hep-lat/9303005].

\bibitem{Furman:1994ky} 
  V.~Furman and Y.~Shamir,
  Nucl.\ Phys.\ B {\bf 439}, 54 (1995)
  doi:10.1016/0550-3213(95)00031-M
  [hep-lat/9405004].

\bibitem{Luscher:1998du} 
  M.~L\"uscher,
  Nucl.\ Phys.\ B {\bf 549}, 295 (1999)
  doi:10.1016/S0550-3213(99)00115-7
  [hep-lat/9811032].

\bibitem{Luscher:1999un} 
  M.~L\"uscher,
  Nucl.\ Phys.\ B {\bf 568}, 162 (2000)
  doi:10.1016/S0550-3213(99)00731-2
  [hep-lat/9904009].

\bibitem{Kikukawa:2000kd} 
  Y.~Kikukawa and Y.~Nakayama,
  Nucl.\ Phys.\ B {\bf 597}, 519 (2001)
  doi:10.1016/S0550-3213(00)00714-8
  [hep-lat/0005015].

\bibitem{Kikukawa:2001mw} 
  Y.~Kikukawa,
  Phys.\ Rev.\ D {\bf 65}, 074504 (2002)
  doi:10.1103/PhysRevD.65.074504
  [hep-lat/0105032].

\bibitem{Kadoh:2007xb} 
  D.~Kadoh and Y.~Kikukawa,
  JHEP {\bf 0805}, 095 (2008)
  Erratum: [JHEP {\bf 1103}, 095 (2011)]
  doi:10.1088/1126-6708/2008/05/095, 10.1007/JHEP03(2011)095
  [arXiv:0709.3658 [hep-lat]].

\bibitem{'tHooft:1976up} 
  G.~'t Hooft,
  Phys.\ Rev.\ Lett.\  {\bf 37}, 8 (1976).
  doi:10.1103/PhysRevLett.37.8

\bibitem{Fujikawa:1983bg} 
  K.~Fujikawa,
  Phys.\ Rev.\ D {\bf 29}, 285 (1984).
  doi:10.1103/PhysRevD.29.285

\bibitem{Kitano:2004sv} 
  R.~Kitano and I.~Low,
  Phys.\ Rev.\ D {\bf 71}, 023510 (2005)
  doi:10.1103/PhysRevD.71.023510
  [hep-ph/0411133].

\bibitem{Kaplan:2009ag} 
  D.~E.~Kaplan, M.~A.~Luty and K.~M.~Zurek,
  Phys.\ Rev.\ D {\bf 79}, 115016 (2009)
  doi:10.1103/PhysRevD.79.115016
  [arXiv:0901.4117 [hep-ph]].

\bibitem{AlvarezGaume:1983cs} 
  L.~Alvarez-Gaum\'e and P.~H.~Ginsparg,
  Nucl.\ Phys.\ B {\bf 243}, 449 (1984).
  doi:10.1016/0550-3213(84)90487-5

\bibitem{Narayanan:2006rf} 
  R.~Narayanan and H.~Neuberger,
  JHEP {\bf 0603}, 064 (2006)
  doi:10.1088/1126-6708/2006/03/064
  [hep-th/0601210].

\bibitem{Luscher:2009eq} 
  M.~L\"uscher,
  Commun.\ Math.\ Phys.\  {\bf 293}, 899 (2010)
  doi:10.1007/s00220-009-0953-7
  [arXiv:0907.5491 [hep-lat]].

\bibitem{Luscher:2010iy} 
  M.~L\"uscher,
  JHEP {\bf 1008}, 071 (2010)
  Erratum: [JHEP {\bf 1403}, 092 (2014)]
  doi:10.1007/JHEP08(2010)071, 10.1007/JHEP03(2014)092
  [arXiv:1006.4518 [hep-lat]].

\bibitem{Luscher:2011bx} 
  M.~L\"uscher and P.~Weisz,
  JHEP {\bf 1102}, 051 (2011)
  doi:10.1007/JHEP02(2011)051
  [arXiv:1101.0963 [hep-th]].

\bibitem{Ginsparg:1981bj} 
  P.~H.~Ginsparg and K.~G.~Wilson,
  Phys.\ Rev.\ D {\bf 25}, 2649 (1982).
  doi:10.1103/PhysRevD.25.2649

\bibitem{Luscher:1998pqa} 
  M.~L\"uscher,
  Phys.\ Lett.\ B {\bf 428}, 342 (1998)
  doi:10.1016/S0370-2693(98)00423-7
  [hep-lat/9802011].

\bibitem{Niedermayer:1998bi} 
  F.~Niedermayer,
  Nucl.\ Phys.\ Proc.\ Suppl.\  {\bf 73}, 105 (1999)
  doi:10.1016/S0920-5632(99)85011-7
  [hep-lat/9810026].

\bibitem{Kikukawa:1998pd} 
  Y.~Kikukawa and A.~Yamada,
  Phys.\ Lett.\ B {\bf 448}, 265 (1999)
  doi:10.1016/S0370-2693(99)00021-0
  [hep-lat/9806013].

\bibitem{Fujikawa:1998if} 
  K.~Fujikawa,
  Nucl.\ Phys.\ B {\bf 546}, 480 (1999)
  doi:10.1016/S0550-3213(99)00042-5
  [hep-th/9811235].

\bibitem{Adams:1998eg} 
  D.~H.~Adams,
  Annals Phys.\  {\bf 296}, 131 (2002)
  doi:10.1006/aphy.2001.6209
  [hep-lat/9812003].

\bibitem{Suzuki:1998yz} 
  H.~Suzuki,
  Prog.\ Theor.\ Phys.\  {\bf 102}, 141 (1999)
  doi:10.1143/PTP.102.141
  [hep-th/9812019].

\bibitem{Nussinov:1985xr}
  S.~Nussinov,
  Phys.\ Lett.\ B {\bf 165} (1985) 55.
  doi:10.1016/0370-2693(85)90689-6

\bibitem{Barr:1990ca} 
  S.~M.~Barr, R.~S.~Chivukula and E.~Farhi,
  Phys.\ Lett.\ B {\bf 241}, 387 (1990).
  doi:10.1016/0370-2693(90)91661-T

\bibitem{Kaplan:1991ah}
  D.~B.~Kaplan,
  Phys.\ Rev.\ Lett.\  {\bf 68} (1992) 741.
  doi:10.1103/PhysRevLett.68.741

\bibitem{Kuzmin:1996he}
  V.~A.~Kuzmin,
  Phys.\ Part.\ Nucl.\  {\bf 29} (1998) 257
   [Fiz.\ Elem.\ Chast.\ Atom.\ Yadra {\bf 29} (1998) 637]
   [Phys.\ Atom.\ Nucl.\  {\bf 61} (1998) 1107]
  doi:10.1134/1.953070
  [hep-ph/9701269].

\bibitem{Hooper:2004dc}
  D.~Hooper, J.~March-Russell and S.~M.~West,
  Phys.\ Lett.\ B {\bf 605} (2005) 228
  doi:10.1016/j.physletb.2004.11.047
  [hep-ph/0410114].

\bibitem{Ade:2015xua}
  P.~A.~R.~Ade {\it et al.} [Planck Collaboration],
  arXiv:1502.01589 [astro-ph.CO].

\bibitem{Kuzmin:1985mm}
  V.~A.~Kuzmin, V.~A.~Rubakov and M.~E.~Shaposhnikov,
  Phys.\ Lett.\ B {\bf 155} (1985) 36.
  doi:10.1016/0370-2693(85)91028-7

\bibitem{Shaposhnikov:1986jp}
  M.~E.~Shaposhnikov,
  JETP Lett.\  {\bf 44} (1986) 465
   [Pisma Zh.\ Eksp.\ Teor.\ Fiz.\  {\bf 44} (1986) 364].

\bibitem{Shaposhnikov:1987tw}
  M.~E.~Shaposhnikov,
  Nucl.\ Phys.\ B {\bf 287} (1987) 757.
  doi:10.1016/0550-3213(87)90127-1

\bibitem{Cohen:1993nk}
  A.~G.~Cohen, D.~B.~Kaplan and A.~E.~Nelson,
  Ann.\ Rev.\ Nucl.\ Part.\ Sci.\  {\bf 43} (1993) 27
  doi:10.1146/annurev.ns.43.120193.000331
  [hep-ph/9302210].

\bibitem{Fukugita:1986hr}
  M.~Fukugita and T.~Yanagida,
  Phys.\ Lett.\ B {\bf 174} (1986) 45.
  doi:10.1016/0370-2693(86)91126-3

\bibitem{Harvey:1990qw} 
  J.~A.~Harvey and M.~S.~Turner,
  Phys.\ Rev.\ D {\bf 42}, 3344 (1990).
  doi:10.1103/PhysRevD.42.3344

\bibitem{Weinberg:2008zzc} 
  S.~Weinberg,
  ``Cosmology,''
  Oxford, UK: Oxford Univ. Pr. (2008) 593 p

\bibitem{Aad:2015pla} 
  G.~Aad {\it et al.} [ATLAS Collaboration],
  JHEP {\bf 1511}, 206 (2015)
  doi:10.1007/JHEP11(2015)206
  [arXiv:1509.00672 [hep-ex]].

\bibitem{CMS:2016rfr} 
  CMS Collaboration [CMS Collaboration],
  CMS-PAS-HIG-16-016.

\bibitem{Crewther:1979pi}
  R.~J.~Crewther, P.~Di Vecchia, G.~Veneziano and E.~Witten,
  Phys.\ Lett.\ B {\bf 88} (1979) 123
   Erratum: [Phys.\ Lett.\ B {\bf 91} (1980) 487].
  doi:10.1016/0370-2693(80)91025-4, 10.1016/0370-2693(79)90128-X

\bibitem{Patt:2006fw}
  B.~Patt and F.~Wilczek,
  hep-ph/0605188.

\bibitem{Kim:2006af}
  Y.~G.~Kim and K.~Y.~Lee,
  Phys.\ Rev.\ D {\bf 75} (2007) 115012
  doi:10.1103/PhysRevD.75.115012
  [hep-ph/0611069].

\bibitem{MarchRussell:2008yu}
  J.~March-Russell, S.~M.~West, D.~Cumberbatch and D.~Hooper,
  JHEP {\bf 0807} (2008) 058
  doi:10.1088/1126-6708/2008/07/058
  [arXiv:0801.3440 [hep-ph]].

\bibitem{Kim:2008pp}
  Y.~G.~Kim, K.~Y.~Lee and S.~Shin,
  JHEP {\bf 0805} (2008) 100
  doi:10.1088/1126-6708/2008/05/100
  [arXiv:0803.2932 [hep-ph]].

\bibitem{Kanemura:2010sh}
  S.~Kanemura, S.~Matsumoto, T.~Nabeshima and N.~Okada,
  Phys.\ Rev.\ D {\bf 82} (2010) 055026
  doi:10.1103/PhysRevD.82.055026
  [arXiv:1005.5651 [hep-ph]].

\bibitem{Djouadi:2011aa}
  A.~Djouadi, O.~Lebedev, Y.~Mambrini and J.~Quevillon,
  Phys.\ Lett.\ B {\bf 709} (2012) 65
  doi:10.1016/j.physletb.2012.01.062
  [arXiv:1112.3299 [hep-ph]].

\bibitem{Low:2011kp}
  I.~Low, P.~Schwaller, G.~Shaughnessy and C.~E.~M.~Wagner,
  Phys.\ Rev.\ D {\bf 85} (2012) 015009
  doi:10.1103/PhysRevD.85.015009
  [arXiv:1110.4405 [hep-ph]].

\bibitem{Englert:2011aa}
  C.~Englert, T.~Plehn, M.~Rauch, D.~Zerwas and P.~M.~Zerwas,
  Phys.\ Lett.\ B {\bf 707} (2012) 512
  doi:10.1016/j.physletb.2011.12.067
  [arXiv:1112.3007 [hep-ph]].

\bibitem{Mohapatra:1980qe}
  R.~N.~Mohapatra and R.~E.~Marshak,
  Phys.\ Rev.\ Lett.\  {\bf 44} (1980) 1316
   Erratum: [Phys.\ Rev.\ Lett.\  {\bf 44} (1980) 1643].
  doi:10.1103/PhysRevLett.44.1316

\bibitem{Wetterich:1981bx}
  C.~Wetterich,
  Nucl.\ Phys.\ B {\bf 187} (1981) 343.
  doi:10.1016/0550-3213(81)90279-0

\bibitem{Buchmuller:1991ce}
  W.~Buchm\"uller, C.~Greub and P.~Minkowski,
  Phys.\ Lett.\ B {\bf 267} (1991) 395.
  doi:10.1016/0370-2693(91)90952-M

\bibitem{Khalil:2006yi}
  S.~Khalil,
  J.\ Phys.\ G {\bf 35} (2008) 055001
  doi:10.1088/0954-3899/35/5/055001
  [hep-ph/0611205].

\bibitem{Basso:2008iv}
  L.~Basso, A.~Belyaev, S.~Moretti and C.~H.~Shepherd-Themistocleous,
  Phys.\ Rev.\ D {\bf 80} (2009) 055030
  doi:10.1103/PhysRevD.80.055030
  [arXiv:0812.4313 [hep-ph]].

\bibitem{Makino:2016auf} 
  H.~Makino and O.~Morikawa,
  arXiv:1609.08376 [hep-lat].

\end{thebibliography}
\end{document}